
%
%
%
%
%
%
\xdef\chapsym{}
\global\newcount\figno \global\figno=1
\input epsf.tex
\ifx\answ\bigans
\else				

\fi
\def\writefigref#1#2{}
%
%
\def\figPS{Fig.~\chapsym\the\figno\nfigPS}
\def\nfigPS#1#2#3{\xdef#1{Fig.~\chapsym\the\figno}%
\mytopinsert{\@PSfig{#2}{#3}\writefigref{#1}{#3}}%
\global\advance\figno by1%
}%
%
%
\def\figPSmid{Fig.~\chapsym\the\figno\nfigPSmid}
\def\nfigPSmid#1#2#3{\xdef#1{Fig.~\chapsym\the\figno}%
\bigbreak\@PSfig{#2}{#3}\writefigref{#1}{#3}\bigskip%
\global\advance\figno by1%
}%
%
%
\def\figuresPS{\extra{Figures}\global\figno=1%
\def\chapsym{}%
\def\figPS##1##2##3{
\xdef##1{Fig.~\chapsym\the\figno}%
\ \vfill\@PSfig{##2}{##3}\writefigref{##1}{##3}\vskip1in\eject%
\global\advance\figno by1%
}}
%
%
\def\@PSfig#1#2{
\centerline{\epsffile{#1}}
\medskip
\centerline{\hfill
\hbox {\bf Figure \chapsym\the\figno:}
\vtop{\advance\hsize by \shrinkdim\baselineskip=12pt\noindent 
\tenrm #2}
\hfill}}
\def\mytopinsert#1{\insert\topins{\penalty100
\splittopskip=0pt\splitmaxdepth=\maxdimen\floatingpenalty=0
\vbox{#1}\nobreak\bigskip}}
\newdimen\shrinkdim
\ifx\ans\bigans\shrinkdim=-2in\else\shrinkdim=-1.5in\fi
\input epsf.tex
\headline={\ifnum\pageno=1\firstheadline\else
\ifodd\pageno\rightheadline \else\leftheadline\fi\fi}
\def\firstheadline{\hfil}
\def\rightheadline{\hfil}
\def\leftheadline{\hfil}
        \footline={\ifnum\pageno=1\firstfootline\else\otherfootline\fi}
\def\firstfootline{\rm\hss\folio\hss}
\def\otherfootline{\hfil}

\font\tenrm=cmr10
\font\tenit=cmti10
\font\elevenbf=cmbx10 scaled\magstep 1
\font\elevenrm=cmr10 scaled\magstep 1
\font\elevenit=cmti10 scaled\magstep 1

\font\ninerm=cmr9

\nopagenumbers
\line{\hfil }
\vglue 1cm
\hsize=6.0truein
\vsize=8.5truein
\parindent=3pc
\baselineskip=10pt
\noindent{\elevenrm CTP\#2251 \hfil hep-th/9311034}
\bigskip
\centerline{\elevenbf SYMMETRIES OF THE DISSIPATIVE HOFSTADTER MODEL
   \footnote{*}
{\ninerm\baselineskip=11pt  This work was supported in part by DOE grant
\#DE-AC02-76ER03069 and NSF grant \#87-8447. \hfil}}
\vglue 1.0cm
\centerline{\tenrm DENISE E.~FREED}
\baselineskip=13pt
\centerline{\tenit Center for Theoretical Physics}
\baselineskip=12pt
\centerline{\tenit Massachusetts Institute of Technology}
\baselineskip=12pt
\centerline{\tenit Cambridge, MA 02139, U.S.A.}
\vglue 0.8cm
\centerline{\tenrm ABSTRACT}
\vglue 0.3cm
{\rightskip=3pc
 \leftskip=3pc
 \tenrm\baselineskip=12pt
 \noindent
The dissipative Hofstadter model, which describes a particle in 2-D subject
to a periodic potential, uniform magnetic field, and dissipation, is also
related to open string boundary states.  This model exhibits an SL(2,Z)
duality symmetry and hidden reparametrization invariance symmetries.  These
symmetries are useful for finding exact solutions for correlation
functions.
\medskip
\centerline{Talk presented at the Strings '93 conference in
Berkeley, May 1993}
}
\vglue 0.8cm
\line{\elevenbf 1. The Dissipative Hofstadter Model and Open String
Theory\hfil}
\smallskip
\baselineskip=14pt
\elevenrm
The dissipative Hofstadter model describes the quantum mechanics of a
particle confined to two dimensions in a periodic potential and transverse
magnetic field, subject to a dissipative force.  The particle's
coordinates are taken to be $\vec x(t) = \left(x(t), y(t)\right)$, and the
magnetic field is given by $B\hat z$.  Classically, the dissipation is
described by the $-\eta \dot{\vec x}$ term in the equations of motion.  To
treat the dissipation quantum mechanically, we use the Caldeira-Leggett
model.$^1$~   In this model, the Euclidean action is given by
$${\cal S} = S_q+S_\eta+S_V,\eqno(1)$$   
where $S_q$ is the usual action of a particle in a constant magnetic field,
$$S_q=\int_{-T/2}^{T/2}dt \left[{M\over2}\dot{\vec x}^2 +{ieB\over 2c}
\left(\dot x y -\dot y x\right)\right].\eqno(2)$$
$S_\eta$ is a non-local kinetic term that accounts for the friction.
It is given by
$$S_\eta={\eta\over 4\pi}\int_{-T/2}^{T/2}\int_{-\infty}^\infty dt\, dt'
\left({\vec x(t)-\vec x(t')\over t-t'}\right)^2.\eqno(3)$$
$S_V$ is the action due to the potential, which we are taking to be
$$S_V = -\int_{-T/2}^{T/2}\left[V_0\cos\left({2\pi x(t)\over a}\right)
                   +V_0\cos\left({2\pi y(t)\over a}\right)\right].\eqno(4)$$

The friction term, $S_{\eta}$, is obtained by first introducing a bath of
harmonic oscillators which interact linearly with the particle.  The
integration over the oscillators in the resulting functional integral
yields the $S_\eta$ term in the remaining action for the particle.

It is useful to define the dimensionless parameters describing the
friction/unit cell, $\alpha$, and the flux/unit cell, $\beta$.  They are
given by
$$2\pi\alpha = {a^2\eta/\hbar} \qquad{\rm and} \qquad
   2\pi \beta = {e B\over \hbar c} a^2.\eqno(5)$$

The action in Eq.~(1) is also used to obtain the boundary state in open
string theory.$^2$~  This boundary state describes a world sheet with a
boundary, where all the fields on the interior of the world sheet are free
and all the interactions take place on the boundary.$^3$~  Then the potential
corresponds to a tachyon field; the magnetic field corresponds to a guage
field; and the mass term acts as a regulator.  In order to give a solution
to string theory, the theory must be independent of the regulator.  Hence
it must be scale invariant and lie at a critical point.
\vglue 0.6cm
\line{\elevenbf 2. The Theory at $\alpha =1$, $\beta=0$\hfil}
\vglue 0.4cm
When $\alpha=1$ and there is no magnetic field,
the theory is expected to be at a critical point.$^4$~ At this point,
we can use simple algebraic identities to fermionize the theory as
follows:$^{5,6}$
$$e^{ix(t)} = {\psi_L}^\dagger(t)\psi_R(t),\eqno(6)$$
$$\dot x(t) = i\left[{\psi_L}^\dagger(t)\psi_L(t) -
                     {\psi_R}^\dagger(t)\psi_R(t)\right],\eqno(7)$$
and similarly for $y(t)$.  The propagator is then given by
$$\langle{\psi_L}^\dagger(t)\psi_L(0) \rangle
     =\langle{\psi_R}^\dagger(t)\psi_R(0)\rangle = {i\over t} \qquad
   {\rm and} \qquad
   \langle{\psi_L}^\dagger(t)\psi_R(0) \rangle =0.\eqno(8)$$
The theory is bilinear in $\psi$, so it can be solved exactly.  We find that
$$\langle \dot x(t_1) \dot x(t_2)\rangle = -\mu{2\over(t_1-t_2)^2},\eqno(9)$$
where $\mu$ is known as the mobility; and all other $m$-point functions of
the $\dot x(t)$'s are contact terms.  That means they are zero unless
at least two points are coincident.

The problem with this treatment is that fermionization is only fine for
large-time behavior, since the derivation ignored how the short distance
behavior was regulated.  One would hope that for the contact terms to be
well-defined and independent of the regulator, the symmetries and possibly
also the large-time behavior of the system are enough to determine them.
\vglue 0.6cm
\line{\elevenbf 3. Duality Symmetry \hfil}
\vglue 0.4cm
The first symmetry we expect this system to have is a duality symmetry in
$\alpha$ and $\beta$.$^5$~ In

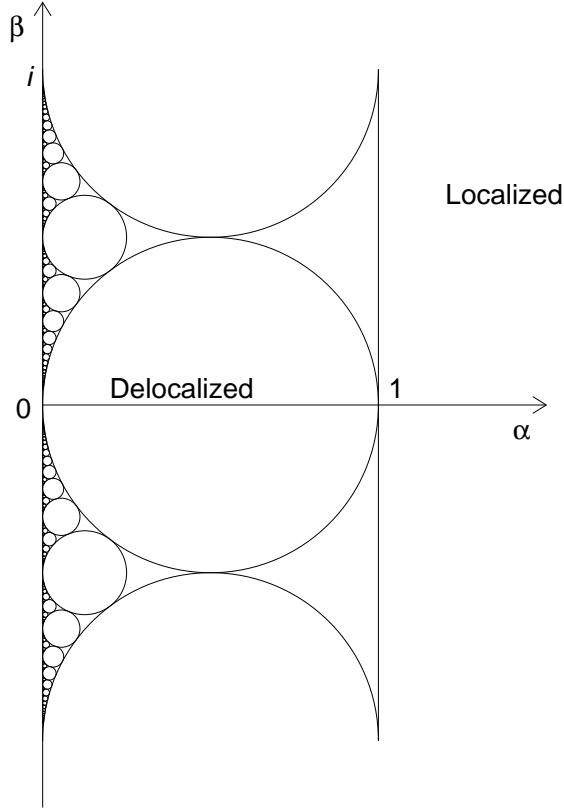
\figPS\figone{figone.epsf}{Phase diagram as a function of friction,
$\alpha$, and magnetic flux, $\beta$.$^5$},
the approximate phase diagram for this system
shows the expected transitions between localized states
(for $\alpha > 1$), delocalized states (in the interiors of the circles),
and unknown states (in the triangular regions between the circles).
The diagram exhibits an
${\rm SL}(2,Z)$ symmetry.  This means that if the theory at one value of
$\alpha$ and $\beta$ is critical, then, for any other value of $\alpha$ and
$\beta$ that is related to the first by an ${\rm SL}(2,Z)$ transformation,
it will also be critical.  In addition, this symmetry relates theories at
different values of flux and friction, so if the correlation functions are
known at one value of $\alpha$ and $\beta$,  then there are simple
transformations we can use to obtain them at the other values of $\alpha$
and $\beta$ related by the ${\rm SL}(2,Z)$ symmetry.  For example, if we know
the correlation function
${\langle \dot{\tilde x}^{\mu_1}(k_1)\ldots \dot{\tilde
x}^{\mu_m}(k_m)\rangle}_0$
at the point $\alpha = 1$, $\beta =0$, then we can obtain all the
correlation functions
${\langle \dot{\tilde x}^{\mu_1}(k_1)\ldots \dot{\tilde
x}^{\mu_m}(k_m)\rangle}_\beta$
at the other multi-critical points on the large circle centered at $\alpha =
1/2$, $\beta=0$, as follows:$^6$
$${\langle \dot{\tilde x}^{\mu_1}(k_1)\ldots
\dot{\tilde x}^{\mu_m}(k_m)\rangle}_\beta =
\left[\prod_{i=1}^m r^{\mu_i x}(k_i)+\prod_{i=1}^m r^{\mu_i y}(k_i)\right]
 {\langle \dot{\tilde x}^{\mu_1}(k_1)\ldots
   \dot{\tilde x}^{\mu_m}(k_m)\rangle}_0,\eqno(10)$$  
where $r^{\mu\nu}$ is given by
$$r^{\mu\nu}(k) = \delta^{\mu\nu} - {\beta \over \alpha} {\rm sign}(k)
\epsilon^{\mu\nu}.\eqno(11)$$

It is not difficult to do calculations to $O(V_0^2)$, and, to this order,
direct calculations show that Eq.~(10) is exact.  They also show that
for $\beta\ne0$, not all the $m$-point
functions are contact terms.$^7$~  This implies that the duality symmetry in
Eq.~(10) relates contact terms at $\beta=0$ to functions that are
non-zero at large times.  We conclude from this that the duality symmetry
should be a guiding principle in finding the contact terms.
\vglue 0.6cm
\line{\elevenbf 4. Reparametrization Invariance \hfil}
\vglue 0.4cm
The second symmetry the system should have comes from the
reparametrization invariance of open string theory.  This symmetry implies
that the generating function for critical dissipative quantum systems
should satisfy ``hidden" reparametrization invariance Ward identities.$^8$~
The generating function, $W[\vec J]$, is given by
$$e^{W[\vec J]}=
  \int D \vec x(t) \exp\left[-{1\over \hbar}(S_q+S_\eta+S_V)\right]
            \exp\left[-{1\over \hbar}\int\vec J \cdot \dot{\vec x} dt\right].
\eqno(12)$$
 For $n\ge0$, the Ward identity is
$$\sum_{m=1}^{n-1}\left[{1\over2}{\partial W\over\partial \vec J_{-m}} \cdot
            {\partial W\over\partial \vec J_{m-n}} -
            {\partial^2 W\over\partial\vec J_{-m}\cdot \vec J_{m-n}}
      \right]
      +\sum_{m = -\infty \atop m\ne0}^\infty m \vec J_m \cdot
            {\partial W\over\partial \vec J_{m-n}} =0,\eqno(13)$$
and there is a similar equation for $n<0$.
This identity is satisfied at all orders in $V_0$ when $\alpha /
(\alpha^2 + \beta^2)=1$ and $\beta/\alpha \in Z$.$^9$~  The
equations with $n=0, \pm1$ imply that, inside the correlation functions,
the $\dot{\vec x}(t_i)$ should
transform as dimension-one operators under ${\rm SL} (2,R)$
transformations of time.  The only other independent equation is
with $n=2$.  One problem with the Ward identities is that they do not give
enough information for solving for the correlation functions.  However, they
do say that any correlation function at $\beta=0$, $\alpha
=1$ must be ${\rm SL}(2,R)$ covariant.  We can apply the duality
transformation to this correlation function to obtain one at another value
of $\beta$.  This new correlation function must once again exhibit the
${\rm SL}(2,R)$ symmetry.  This should give a lot of information about the
form of the correlation functions.

One difficulty with this reasoning is that the regulator used to derive the
duality transformation is not the same one used to prove the Ward
identities.  A rigorous derivation of Eq.~(10) using the regulator
satisfying the Ward identities gives a slightly weaker version of the
transformation.
\vglue 0.6cm
\line{\elevenbf 5. Results for Correlation Functions \hfil}
\vglue 0.4cm
If we carefully repeat the derivation of fermionization using the regulator
that satisfies the Ward identity, then we can obtain additional symmetries
and properties of the correlation functions to {\it all} orders in $V_0$.
We find that, when $\alpha/(\alpha^2 + \beta^2)=1$ and $\beta/\alpha\in Z$,
the $m$-point functions are given by$^6$
$${\langle \dot{\tilde x}^{\mu_1}(k_1)\ldots
    \dot{\tilde x}^{\mu_m}(k_m)\rangle}_\beta =
\left[\prod_{j=1}^m r^{\mu_j x}(k_j)+\prod_{j=1}^m r^{\mu_j y}(k_j)\right]
  F(\vec k, \beta).\eqno(14)$$
This is the form of the correlation functions predicted by the duality
transformation in Eq.~(10).  The function
$F(\vec k,\beta) = \vec a(\vec k) \cdot \vec k$ has the following
properites: It is finite as the cutoff goes to zero.  It is piecewise linear
in $\vec k$, which suggests a connection with the Duistermaat-Heckman
theorem;$^{10}$ and it is homogeneous in $\vec k$.  In real space, this last
property gives a new symmetry under non-one-to-one reparametrizations of
time. It tells what happens when $z \to z^n$, where $z = e^{2\pi i k t}$.  In
addition, we find that $F(\vec k, \beta) \to 0$ if $k_i = 0$; $F$ is
symmetric under $k_i \leftrightarrow k_j$, and also $\vec k \leftrightarrow
- \vec k$; and $F$ is continuous.  These results restrict the $m$-point
functions to lie in a finite-dimensional linear space.

We conjecture that these results, combined with the reparametrization
invariance Ward identites, determine all the $m$-point functions when
$\alpha/(\alpha^2 + \beta^2) =1$ and $\beta/\alpha \in Z$.  We have
found that they do give exact solutions for the two, four and six-point
functions, and also for any correlation function with special conditions on
the $k_i's$.$^{11}$~  For example, the 4-point function is proportional to
${\rm min} \left(|k_1|, |k_2|, |k_3|, |k_4|\right)$.  However, it is still
an open question whether these symmetries are enough to determine any
arbitrary correlation function.  In any case, it does appear that these
symmetries, combined with the long-time behavior, are enough to obtain
exact solutions for all the correlation functions.
\vglue 0.6cm
\line{\elevenbf 6. Acknowledgements \hfil}
\vglue 0.4cm
I would like to thank the organizers for arranging such an
interesting and stimulating conference.
\vglue 0.6cm
\line{\elevenbf 7. References \hfil}
\vglue 0.4cm
%
%
\item{1.} A.~O.~Caldeira and A.~J.~Leggett, {\elevenit Physica}
{\elevenbf 121A}(1983) 587;
{\elevenit Phys. Rev. Lett.} {\elevenbf 46} (1981) 211;
{\elevenit Ann. of Phys.} {\elevenbf 149} (1983) 374.
\item{2.} C.~G.~Callan, L.~Thorlacius,  {\elevenit Nucl. Phys.}
{\elevenbf B329} (1990) 117.
\item{3.}C.~G.~Callan, C.~Lovelace, C.~R.~Nappi, and S.~A.~Yost,
{\elevenit Nucl. Phys.} {\elevenbf B293} (1987) 83;
{\elevenit Nucl. Phys.} {\elevenbf B308} (1988) 221.
\item{4.} A.~Schmid, {\elevenit Phys. Rev. Lett.} {\elevenbf 51} (1983) 1506;
 F.~Guinea, V.~Hakim and A.~Muramatsu, {\elevenit Phys. Rev. Lett.}
{\elevenbf 54} (1985) 263;
M.~P.~A.~Fisher and W.~Zwerger, {\elevenit Phys. Rev.}
{\elevenbf B32} (1985) 6190.
\item{5.}C.~G.~Callan and D.~Freed, {\elevenit Nucl. Phys}
{\elevenbf B374}(1992)543.
\item{6.}D.~E.~Freed, CTP\#2170, to appear in {\elevenit Nucl. Phys.}
{\elevenbf B}.
\item{7.}C.~G.~Callan, A.~G.~Felce and D.~E.~Freed, {\elevenit Nucl. Phys.}
{\elevenbf B392} (1993) 551.
\item{8.}C.~G.~Callan, L.~Thorlacius, {\elevenit Nucl.  Phys.}
{\elevenbf B319} (1989) 133.
\item{9.}D.~E.~Freed, CTP\#2241, submitted to {\elevenit Nucl. Phys.}
{\elevenbf B}.
\item{10.}J.~J.~Duistermaat, G.~J.~Heckman, {\elevenit Invent. Math}
{\elevenbf 69} (1982)259.
\item{11.}D.~E.~Freed, in preparation.
\vfil
\eject
\bye